# Optical Target Tracking by Scheduled Range Measurements

**Mohammad Hossein Ferdowsi,[a] Ebrahim Sabzikar[a]**

[a] Control Group, Electrical Eng. Department, Malek-e-Ashtar University of Technology, Lavizan, Tehran, Iran, Post Box 15875-1774

**Abstract.** In this paper, optical target tracking, by regular target bearing measurements and target range in a lower and scheduled measurement rate is considered. Variance of the target range estimation error is used as scheduling criterion. For this purpose, target dynamic state vector in modified spherical coordinates is stated in such a way that all target states be decoupled from range-related target state. Target state dynamic equations in modified spherical coordinates for nearly constant velocity, nearly constant acceleration and coordinated turn rate kinematic models, are analytically derived. For resulted state dynamic equations, a UKF-IMM filter with range measurement scheduling is utilized as a tracking filter. It is shown that target states are estimated properly and applied filter has high performance in maneuvering target tracking.

## 1 Introduction

The subject of this paper is target state estimation using one stationary optical tracker. Target bearings are measured by tracking platform. Target range can be measured using active range finder or it can be estimated using triangulation method by two or more passive stationary tracking sensor at specified locations[1] or using one sensor with proper maneuvers that make the target range observable[2]. In practical situations when there is incomplete passive measurements like target motion analysis with restricted own-ship motion, scheduling active measurements can be used for generating target observability conditions. In the case of only one stationary tracking sensor, utilizing active range measurement mechanism such as laser range finder is necessary, however tactical considerations limit continuous active range measurements. A solution for this problem is range measurement scheduling, but measurement scheduling requires a criterion. The most appropriate criterion for range measurement scheduling is Posterior Cramer-Rao lower bound (PCRLB) of range estimation error[3] **which provides a measure of the optimal achievable accuracy of target state estimation.** However, calculation of PCRLB is problematic, and except for special cases[4], no closed form solution is available. Moreover, Monte-Carlo based methods for PCRLB calculation are time consuming and not suitable for real time implementation. If the estimation of target states by tracking filter is consistent[5], state estimation error covariance is a suitable PCRLB substitute as the scheduling criterion. According to the criterion, if range estimation variance value exceeds a predetermined threshold, range has to be measured.

For optical target tracking in Cartesian coordinates, measured bearings are nonlinear functions of target states, so that for estimating target states, nonlinear Kalman filter should be used[5]. In the case of moderate nonlinearity in the target states dynamic equations or measurement equations, the Extended Kalman Filter (EKF) can be used. In the case of high degree of nonlinearity in state dynamic equations, other filtering methods like Unscented Kalman Filter (UKF) or Particle Filter (PF) instead of EKF must be used[6].

With one stationary target tracking sensor, if the target range measurement is not available, target range estimation will diverge. If the target states are expressed in Cartesian or Spherical coordinates (SC), by range estimation divergence, all other estimated states will diverge too; because there is a correlation between the unobservable target range and the other target states[7].

In the case of states divergence, estimation of tracking filter is not consistent and state estimation error covariance is not suitable for scheduling. In order to solve the above problem, the target states can be expressed in Modified Spherical Coordinate (MSC) system[8,9]. In this way, inverted range is a target state and dynamic equations of all other target states are independent of this state. Therefore with continuous bearing measurements and range measurement scheduling using range estimation error variance criterion, all the target states can be estimated without considerable divergence.

In this paper a method for scheduling active range measurements in maneuvering target tracking using one stationary optical tracking sensor is presented. In order to schedule the target range measurements, the variance of target range estimation error is considered as scheduling criterion. In the case of expressing target states in MSC, UKF based tracker can produce consistent estimations, therefore this tracker is chosen for implementation. Target maneuverability affects process noise in the target state dynamic equations. For compensating the variation of process noise and considering different maneuvering target models, Interactive Multiple Model (IMM) filter with three different target kinematic models is considered[5] . Kinematic models consist of Nearly Constant Velocity (NCV), Nearly Constant Acceleration (NCA) and Coordinated Turn (CT) models. For Cartesian coordinate frame, mentioned kinematic models are simple and can be derived straightforwardly. However, in SC and MSC systems, NCA and CT models are more complicated and derivation of these kinematic models has not been addressed in previous papers. Therefore target state dynamic equations for NCA and CT kinematic models in MSC are derived such that inverted range state only appears as a process noise in the dynamic equations of the other target states.

Contributions of this paper are:

1. Analytical derivation of the target state dynamic equations for NCV, NCA and CT kinematic models in MSC. Note that in previous works, target state dynamic equations in MSC have been derived only for NCV model by substitution of MSC state variables in target state dynamic equations of Cartesian coordinate frame[9].

2. Presenting a new range measurement scheduling method for maneuvering target tracking.

Performance of the proposed IMM filter is demonstrated by simulation. This paper is organized as follows. Section 2 presents target state dynamic equations in MSC. Section 3 discusses implementation of the tracking filter in MSC. In section 4, EKF and UKF are explained. Section 5 presents IMM filtering technique. In section 6, implementation of range measurements scheduling is described. Section 7 presents simulation results and finally conclusion is summarized in section 8.

## 2 Target state dynamic equations in modified spherical coordinates

Target position relative to the tracking sensor in Cartesian coordinate frame is given as x, y and z. On the other hand, target position relative to the sensor in spherical coordinate frame is given as azimuth angle ($\psi$), elevation angle ($\theta$), and range ($r$). Moreover, in modified spherical coordinates, target position is expressed as $s = \frac{1}{r}, \theta$ and $\psi$. Sensor measurements are azimuth ($\psi_m$), elevation ($\theta_m$) and range ($r_m$) as shown in Fig. 1 and are related to target position as given in (1).

$$\begin{cases} \psi_m = \psi + n_\psi = \operatorname{atan}\left(\frac{y}{x}\right) + n_\psi \\ \theta_m = \theta + n_\theta = \operatorname{atan}\left(\frac{z}{\sqrt{x^2+y^2}}\right) + n_\theta \\ r_m = r + n_r = \frac{1}{s} + n_r = \sqrt{x^2 + y^2 + z^2} + n_r \end{cases}, \quad (1)$$

In (1), $n_\psi, n_\theta$ and $n_r$ are measurement noise in target bearing angles $\psi$ and $\theta$ and range r respectively. Target kinematic models, i.e. NCV, NCA and CT are derived from target motion dynamic equations. In these models, second- or third-order derivatives of the target position are not zero, but a zero-mean random process. In Ref. 5, the state dynamic equations of these kinematic models are developed in Cartesian coordinate frame. In subsequent subsections, the state dynamic equations of kinematic models in MSC are presented.

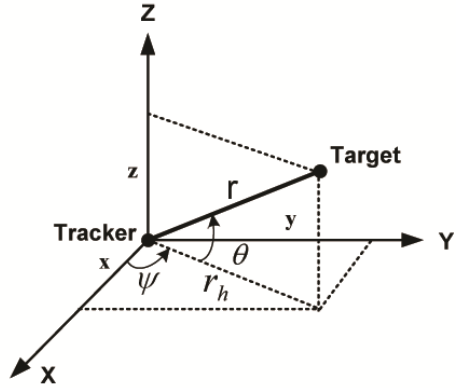

**Fig. 1** Definition of sensor measured angles.

*2.1 Nearly constant velocity target model in MSC*

For nearly constant velocity model in modified spherical coordinates, target states are selected as $\omega = \dot\psi \cos\theta$, $s = \frac{1}{r}$, $\tau = \frac{\dot r}{r}$, $\dot\theta$, $\theta$ and $\psi$. Target state vector is defined as (2).

$$X_{CV}^{ms} = [\omega \quad \dot\theta \quad \tau \quad \psi \quad \theta \quad s]^T = [x_1 \quad x_2 \quad x_3 \quad x_4 \quad x_5 \quad x_6]^T, \qquad (2)$$

Dynamic equations of target states are obtained as (3).

$$\begin{cases} \frac{d\omega}{dt} = -2\tau\omega + \dot{\theta}\omega\,\tan\theta + s\,w_y^s \\ \frac{d\dot{\theta}}{dt} = -\omega^2\tan\theta - 2\dot{\theta}\tau + s\,w_z^s \\ \frac{d\tau}{dt} = \dot{\theta}^2 + \omega^2 - \tau^2 + s\,w_x^s \\ \frac{d\psi}{dt} = \frac{\omega}{\cos\theta} \\ \frac{d\theta}{dt} = \dot{\theta} \\ \frac{ds}{dt} = -\tau\,s \end{cases}, \quad (3)$$

Derivation of these equations is given in appendix A. In (3), $w_x^s$, $w_y^s$ and $w_z^s$ are elements of acceleration process noise vector stated in spherical coordinate frame. In MSC, target state dynamic equations are nonlinear and target range is a nonlinear function of state $s$. An important point of these equations is that the state $s$ does not appear in the dynamic equations of the other states and only appears as a gain in the process noise.

## 2.2 Nearly constant acceleration target model in MSC

For nearly constant acceleration target model, state vector is defined as (4).

$$X_{CA}^{ms} = \begin{bmatrix} \omega & \dot{\theta} & \tau & \psi & \theta & s & \sigma_x^c & \sigma_y^c & \sigma_z^c \end{bmatrix}^T = \begin{bmatrix} x_1 & x_2 & x_3 & x_4 & x_5 & x_6 & x_7 & x_8 & x_9 \end{bmatrix}^T, \quad (4)$$

Three new states $\sigma_x^c$, $\sigma_y^c$ and $\sigma_z^c$ are defined in appendix B.

If nearly constant acceleration target model in Cartesian coordinate frame is transformed to MSC, the equations (5) for target state dynamic equations in MSC are obtained:

$$\begin{cases} \frac{d\omega}{dt} = -2\tau\omega + \dot{\theta}\omega \tan\theta - \sin\psi\, \sigma_x^C + \cos\psi\, \sigma_y^C \\ \frac{d\dot{\theta}}{dt} = -\omega^2 \tan\theta - 2\dot{\theta}\tau - \sin\theta\cos\psi\, \sigma_x^C - \sin\theta\sin\psi\, \sigma_y^C + \cos\theta\, \sigma_z^C \\ \frac{d\tau}{dt} = \dot{\theta}^2 + \omega^2 - \tau^2 + \cos\theta\cos\psi\, \sigma_x^C + \cos\theta\sin\psi\, \sigma_y^C + \sin\theta\, \sigma_z^C \\ \frac{d\psi}{dt} = \frac{\omega}{\cos\theta} \\ \frac{d\theta}{dt} = \dot{\theta} \\ \frac{ds}{dt} = -\tau s \\ \frac{d\sigma_x^C}{dt} = -\tau\sigma_x^C + s w_x \\ \frac{d\sigma_y^C}{dt} = -\tau\sigma_y^C + s w_y \\ \frac{d\sigma_z^C}{dt} = -\tau\sigma_z^C + s w_z \end{cases} \quad , \tag{5}$$

Derivation of the equations (5) is given in appendix B. In (5), $w_x$, $w_y$ and $w_z$ are jerk process noise in x, y and z axis of Cartesian coordinate frame respectively. In these equations, similar to NCV model, the state $s$ does not appear in dynamic equations of the other states and only appears as a gain in the process noise.

*2.3 Coordinated turn target model in MSC*

For coordinated turn target model, target state vector is defined as (6).

$$X_{CT}^{ms} = [\omega \quad \dot{\theta} \quad \tau \quad \psi \quad \theta \quad s \quad \omega_T]^T = [x_1 \quad x_2 \quad x_3 \quad x_4 \quad x_5 \quad x_6 \quad x_7]^T, \tag{6}$$

In (6) $\omega_T$ is the target turn rate in x-y plane in Cartesian coordinate frame. If CT model in Cartesian coordinate frame is transformed to MSC, target state dynamic equations in MSC are obtained as (7).

$$\begin{cases} \frac{d\omega}{dt} = \tau\omega_T \cos\theta - 2\omega\tau - \dot{\theta}\omega_T \sin\theta + \dot{\theta}\,\omega \tan\theta + s\,w_y^S \\ \frac{d\dot{\theta}}{dt} = -\omega^2 \tan\theta + \omega_T \sin\theta\,\omega - 2\dot{\theta}\tau + s\,w_z^S \\ \frac{d\tau}{dt} = \dot{\theta}^2 + \omega^2 - \tau^2 - \omega_T \cos\theta\,\omega + s\,w_x^S \\ \frac{d\psi}{dt} = \frac{\omega}{\cos\theta} \\ \frac{d\theta}{dt} = \dot{\theta} \\ \frac{ds}{dt} = -\tau\,s \\ \frac{d\omega_T}{dt} = w_{\omega_T} \end{cases}, \qquad (7)$$

Where in (7), $w_{\omega_T}$ is the turn rate process noise and $w_x^S$, $w_y^S$ and $w_z^S$ are acceleration process noise in x, y and z axis of spherical coordinate frame respectively. Derivation of the above equations is given in appendix C. Similarly in the above equations, the state $s$ does not appear in the dynamic equations of other states except state s and only appears as a gain in the process noise in the first three equations of (7).

## 3  Implementation of tracking filter in MSC

Nonlinear dynamic equations of target states in MSC are specified by stochastic differential equations[10] and are stated as (8).

$$\frac{d}{dt} X^{ms}(t) = \mathbf{f}(X^{ms}(t)) + \mathbf{g}(X^{ms}(t))w, \qquad (8)$$

In (8), **f** and **g** are defined as (9) for NCV model, as (10) for NCA model and as (11) for CT model.

$$\mathbf{f}_{CV}(X_{CV}^{ms}) = \begin{bmatrix} -2x_3 x_1 + x_2 x_1 \tan(x_5) \\ -x_1^2 \tan(x_5) - 2x_2 x_3 \\ x_2^2 + x_1^2 - x_3^2 \\ x_1/\cos(x_5) \\ x_2 \\ -x_3 x_6 \end{bmatrix}, \quad \mathbf{g}_{CV}(X_{CV}^{ms}) = \begin{bmatrix} x_6 & 0 & 0 \\ 0 & x_6 & 0 \\ 0 & 0 & x_6 \\ 0 & 0 & 0 \\ 0 & 0 & 0 \\ 0 & 0 & 0 \end{bmatrix}, \qquad (9)$$

$$\mathbf{f}_{CA}(X_{CA}^{ms}) = \begin{bmatrix} -2x_3x_1 + x_2x_1 \tan x_5 - \sin x_4 \ x_7 + \cos x_4 \ x_8 \\ -x_1^2 \tan(x_5) - 2x_2x_3 - \sin x_5 \cos x_4 \ x_7 - \sin x_5 \sin x_4 \ x_8 + \cos x_5 \ x_9 \\ x_2^2 + x_1^2 - x_3^2 + \cos x_5 \cos x_4 \ x_7 + \cos x_5 \sin x_4 \ x_8 + \sin x_5 \ x_9 \\ x_1/\cos(x_5) \\ x_2 \\ -x_3x_6 \\ -x_3x_7 \\ -x_3x_8 \\ -x_3x_9 \end{bmatrix},$$

$$\mathbf{g}_{CA}(X_{CA}^{ms}) = \begin{bmatrix} 0 & 0 & 0 \\ 0 & 0 & 0 \\ 0 & 0 & 0 \\ 0 & 0 & 0 \\ 0 & 0 & 0 \\ 0 & 0 & 0 \\ x_6 & 0 & 0 \\ 0 & x_6 & 0 \\ 0 & 0 & x_6 \end{bmatrix}, \tag{10}$$

$$\mathbf{f}_{CT}(X_{CT}^{ms}) = \begin{bmatrix} -2x_3x_1 + x_2x_1 \tan x_5 - x_2x_7 \sin x_5 + x_3x_7 \cos x_5 \\ -x_1^2 \tan x_5 - 2x_2x_3 + x_1x_7 \sin x_5 \\ x_2^2 + x_1^2 - x_3^2 - x_1x_7 \cos x_5 \\ x_1/\cos(x_5) \\ x_2 \\ -x_3x_6 \\ 0 \end{bmatrix}, \mathbf{g}_{CT}(X_{CT}^{ms}) = \begin{bmatrix} 0 & x_6 & 0 & 0 \\ 0 & 0 & x_6 & 0 \\ x_6 & 0 & 0 & 0 \\ 0 & 0 & 0 & 0 \\ 0 & 0 & 0 & 0 \\ 0 & 0 & 0 & 0 \\ 0 & 0 & 0 & 1 \end{bmatrix}, \tag{11}$$

Power spectral density matrix of process noises ($w$) in continuous case for NCV, NCA and CT models are computed as (12), (13) and (14) respectively.

$$Q_{CV} = C_c^s Q_w C_c^{sT} = C_c^s \begin{bmatrix} \sigma_x^2 & 0 & 0 \\ 0 & \sigma_y^2 & 0 \\ 0 & 0 & \sigma_z^2 \end{bmatrix} C_c^{sT}, \tag{12}$$

$$Q_{CA} = Q_w = \begin{bmatrix} \sigma_x^2 & 0 & 0 \\ 0 & \sigma_y^2 & 0 \\ 0 & 0 & \sigma_z^2 \end{bmatrix}, \tag{13}$$

$$Q'_{CT} = C_c^s Q_w C_c^{sT} = C_c^s \begin{bmatrix} \sigma_x^2 & 0 & 0 \\ 0 & \sigma_y^2 & 0 \\ 0 & 0 & \sigma_z^2 \end{bmatrix} C_c^{sT}, Q_{CT} = \begin{bmatrix} Q'_{CT} & 0 \\ 0 & q_{\omega_T} \end{bmatrix}, \tag{14}$$

In (12), (13) and (14), $Q_w$ is the power spectral density matrix of process noise in continuous time system and $C_c^s$ is the rotation matrix from Cartesian to spherical Coordinate frame and is defined in Appendix A. Assuming that $T_s$ is frame sampling interval and using first term of stochastic Taylor series of (8), discretized form of state dynamic equations (8) is given as (15).

$$\boldsymbol{X^{ms}}(k+1) \approx \boldsymbol{X^{ms}}(k) + \mathbf{f}(\boldsymbol{X^{ms}}(k))T_s + \boldsymbol{w}(k) = \boldsymbol{F}(\boldsymbol{X^{ms}}(k)) + \boldsymbol{w}(k), \tag{15}$$

In (15), $\boldsymbol{w}(k)$ is discrete process noise. Computation of discrete process noise covariance, $\boldsymbol{Q_d}(k)$, is given in Ref. 5 as (16).

$$\boldsymbol{Q_d}(k) = \int_{t_k}^{t_{k+1}} e^{A(t_{k+1}-\tau)} \mathbf{g}\boldsymbol{Q}\mathbf{g}^{\mathrm{T}} e^{(t_{k+1}-\tau)A^{\mathrm{T}}} d\tau, \tag{16}$$

Where $\boldsymbol{Q}$ stands for $\boldsymbol{Q_{CV}}$, $\boldsymbol{Q_{CA}}$ and $\boldsymbol{Q_{CT}}$, and $\boldsymbol{A}$ is defined as (17).

$$\boldsymbol{A} = \frac{\partial \mathbf{f}(X^{ms})}{\partial X^{ms}}, \tag{17}$$

Expression (16) can be approximated by mid-point rule, so that $\boldsymbol{Q_d}(k)$ can be computed approximately as (18) and (19).

$$\boldsymbol{S} = \mathbf{g}\boldsymbol{Q}\mathbf{g}^{\mathrm{T}}, \tag{18}$$

$$\boldsymbol{Q_d}(k) \approx e^{\frac{A\Delta t}{2}}(\boldsymbol{S}\Delta t)e^{\frac{A^T \Delta t}{2}}, \tag{19}$$

In (19), $\Delta t = t_{k+1} - t_k$ is sampling interval. The relation between measurements and target states is given by (20).

$$\boldsymbol{Z}(k) = \mathbf{h}(\boldsymbol{X^{ms}}(k)) + \boldsymbol{v}(k) = \begin{bmatrix} \psi_m \\ \theta_m \\ r_m \end{bmatrix} = \begin{bmatrix} x_4(k) \\ x_5(k) \\ 1/x_6(k) \end{bmatrix} + \boldsymbol{v}(k), \tag{20}$$

Measurement noise covariance is considered as (21).

$$\boldsymbol{R} = \begin{bmatrix} \sigma_\psi^2 & 0 & 0 \\ 0 & \sigma_\theta^2 & 0 \\ 0 & 0 & \sigma_r^2 \end{bmatrix}, \tag{21}$$

## 4  Extended and Unscented Kalman filters

EKF is the first choice for implementation of the estimators for nonlinear systems. For mean and covariance propagation, EKF needs linearization of state dynamic and measurement equations. However, in dynamic systems with high level of nonlinearity, linearization introduces large errors in propagation of EKF predicted mean and covariance. These errors can sometimes lead to divergence of the filter. In Ref. 5, required recursive equations to implement EKF are given. Because of high level of nonlinearity in MSC target state dynamic equations, EKF cannot produce consistent state estimations and therefore it is not utilized for implementation.

UKF is a derivative-free alternative to EKF and works on the basis of unscented transform for propagating states mean and covariance[6,11,12]. As it is demonstrated in simulation results in section 8, this filter can produce consistent state estimation. UKF is based on two principles. First a nonlinear mapping can simply be applied on a single point and second, a set of distinct deterministic points in the state space can simply be found with the same PDF as the estimated states. Minimum number of chosen points, called sigma points, is selected such that for any nonlinearity to capture accurately the posterior mean and covariance to the second order of Taylor series expansion[6,11]. Nonlinear mapping $F(X)$ is applied to each of the sigma points. Then mean and covariance of the mapped sigma points, which are good estimation for actual mean and covariance, are used for filtering. The UKF algorithm is described as follows.

For each state vector $X$ with dimension $n$ and covariance matrix $P$, desired sigma points with corresponding weights are generated by state vector mean and covariance[12] in the $(k-1)^{th}$ iteration.

$$\begin{cases} X_{k-1|k-1}^i = X_{k-1|k-1} & , \quad w^i = \frac{\lambda}{n+\lambda} \quad i = 0 \\ X_{k-1|k-1}^i = X_{k-1|k-1} + \delta X^i, & w^i = \frac{1}{2(n+\lambda)} \quad i = 1, \dots, 2n \\ \delta X^i = \sqrt{(n+\lambda)P_{k-1|k-1}}\Big|_i & , \quad i = 1, \dots, n \\ \delta X^{(n+i)} = -\sqrt{(n+\lambda)P_{k-1|k-1}}\Big|_i & , \quad i = 1, \dots, n \end{cases} \quad (22)$$

In (22), $\sqrt{P}$ is defined as (23).

$$P = (\sqrt{P})(\sqrt{P})^T, \qquad (23)$$

$\sqrt{P}\big|_i$ is the i$^{th}$ column of $\sqrt{P}$ matrix. In above equations, $n$ is dimension of state vector and $\lambda$ is scaling parameter and is computed by (24).

$$\lambda = \alpha^2(n+\kappa) - n, \qquad (24)$$

Parameter α shows the point distribution around the mean value and usually is taken as $e^{-4} \leq \alpha \leq 1$. $n$ is dimension of state vector and $\kappa$ is chosen as in Ref. 6.

Using nonlinear function $F(X)$, predicted state for each sigma point in the (k)$^{th}$ iteration is computed by (25).

$$X_{k|k-1}^i = F(X_{k-1|k-1}^i), \qquad (25)$$

Predicted states of sigma points are combined as (26) to obtain overall predicted state.

$$\widehat{X}_{k|k-1} = \sum_{i=0}^{2n} w^i X_{k|k-1}^i, \qquad (26)$$

State prediction error covariance is computed by (27).

$$P_{k|k-1} = \sum_{i=0}^{2n} w^i \left(X_{k|k-1}^i - \widehat{X}_{k|k-1}\right)\left(X_{k|k-1}^i - \widehat{X}_{k|k-1}\right)^T + Q_{wsd}(k-1), \qquad (27)$$

In order to update states by measurements, the predicted sigma points $X_{k|k-1}^i$ are propagated through nonlinear function $h(X)$ that is given in measurement equation (20) and is stated as (28).

$$Z_{k|k-1}^i = h(X_{k|k-1}^i) \quad , i = 0, \dots, 2n, \qquad (28)$$

The mean and covariance of predicted observations are computed as (29).

$$\begin{cases} \widehat{Z}_{k|k-1} = \sum_{i=0}^{2n} w^i Z_{k|k-1}^i \\ P_{zz} = \sum_{i=0}^{2n} w^i (Z_{k|k-1}^i - \widehat{Z}_{k|k-1})(Z_{k|k-1}^i - \widehat{Z}_{k|k-1})^T, \\ P_{xz} = \sum_{i=0}^{2n} w^i (X_{k|k-1}^i - \widehat{X}_{k|k-1})(Z_{k|k-1}^i - \widehat{Z}_{k|k-1})^T \end{cases} \quad (29)$$

Kalman filter gain and measurement update equations are computed as (30).

$$\begin{cases} K = P_{xz} P_{zz}^{-1} \\ \widehat{X}_{k|k} = \widehat{X}_{k|k-1} + K(Z_k - \widehat{Z}_{k|k-1}), \\ P_{k|k} = P_{k|k-1} - K P_{zz} K^T \end{cases} \quad (30)$$

## 5  Interacting Multiple Model Filtering

For maneuvering target tracking, tracking filter must be able to adapt its bandwidth according to target maneuverability. There are some adaptive estimation algorithms. A class of adaptive estimation algorithms is multiple model algorithms. These algorithms assume that the system behaves according to one of a finite number of models. The models can differ in noise levels or their structure. Such systems are called hybrid systems. The IMM estimator is a suboptimal hybrid filter and has the ability to estimate the state of a dynamic system with several behavior modes that can switch from one to the other[5,14]. This can be considered as a variable-bandwidth self-adjusting filter and hence is very well suited for tracking maneuvering targets. It consists of a filter for each model, a model probability evaluator, an estimate mixer at the input of the filter, and an estimate combiner at the output of the filter. A derivation and detailed explanation of the IMM filter is given in Ref. 5. In this paper, an IMM filter with three kinematic models is considered. The first model is NCV model that has acceleration process noise with $\sigma = 2\text{m/sec}^2$ standard deviation in x, y and z axis of Cartesian coordinate frame. The second model is a NCA model that has jerk process noise with $\sigma = 15\text{m/sec}^3$ standard deviation in x, y and z axis of Cartesian coordinate frame. The third model is a CT model in x-y plane with turn acceleration noise with $\mathbf{q}_{\omega_T} = 0.05$ rad/sec² standard deviation. The Markov transition probabilities matrix is set as (31).

$$[\mathbf{P}_{ij}] = \begin{bmatrix} 0.990 & 0.005 & 0.005 \\ 0.005 & 0.990 & 0.005 \\ 0.005 & 0.005 & 0.990 \end{bmatrix} \tag{31}$$

All computations for one cycle of the IMM algorithm are similar to Ref. 13.

## 6  Timing and Scheduling of Active Measurements

In measurement scheduling problem, the goal is to find the optimal time distribution of measurements. In Ref. 13, a method is proposed for optimizing the time distribution of measurements for estimating a scalar random variable. In this method, measurement scheduling problem is converted to an optimization problem. In Ref. 15, method of Ref. 13 is extended to discrete time, vectored random process. The goal of the above paper is to determine the time distribution of measurement variance such that the trace of state estimation error covariance of a linear estimator for related random process is minimized. This way, determination of measurement time distribution problem is converted to a nonlinear optimization problem with inequality and equality constraints. This problem is solved by a variant of projected Newton's Method. This method is not suitable for nonlinear estimation problems. In Ref. 3, for solving measurement scheduling problem in a nonlinear stochastic process, use of PCRLB is proposed. In Ref 16, problem of active and passive measurement scheduling for bearings only tracking is solved using PCRLB. The proposed method in this paper for scheduling active measurements in aerial target tracking is similar to the method of Ref. 3 except that, instead of PCRLB, state estimation covariance matrix in MSC is utilized. If the state estimation error covariance of an estimator agrees with actual estimation error covariance (that is estimator is consistent), range estimation error variance (calculated using state estimation error covariance of estimator) can be utilized as a criterion for range measurement scheduling.

Active measurements are used only when scheduling criterion exceeds predetermined threshold. The scheduling criterion is the standard deviation of target range estimation error. Because state vector expressed in MSC includes inverted range ($s$) instead of range ($r$) as given in equations (3), for calculating scheduling criterion, the inverted range variance is converted to range variance. For this purpose, a method similar to unscented transform is utilized as in (32).

$$\begin{cases} s_0 = \hat{s} \quad , \quad w^0 = \frac{\lambda}{1+\lambda} \\ s_i = \hat{s} + \delta s^i \;, \quad w^i = \frac{1}{2(1+\lambda)} \;, \quad \delta s^i = \pm\sqrt{(1+\lambda)\mathbf{P}_s} \quad i = 1,2 \\ r_i = \frac{1}{s_i} \quad i = 0,1,2 \\ \hat{r} = \sum_{i=0}^{2} w^i r_i \\ \mathbf{P}_r = \sum_{i=0}^{2} w^i (r_i - \hat{r})^2 \\ \boldsymbol{\sigma}_r = \sqrt{\mathbf{P}_r} \end{cases}, \qquad (32)$$

In (32), $\hat{s}$ is estimation of state ($s$), $\mathbf{P}_s$ is estimation error variance of state $s$ and parameter $\lambda$ is defined as (24). Resulted $\boldsymbol{\sigma}_r$ is standard deviation of range estimation error. For scheduling active measurements, if the value of $\boldsymbol{\sigma}_r$ exceeds the related threshold, active range measurement is performed once. Threshold value is determined by required estimated range accuracy and allowable active measurement frequency.

## 7  Simulation Results

For performance evaluation of proposed tracking filter and measurement scheduling method, a tracking scenario for a target with complicated maneuvers as a combination of six different maneuver parts is considered as follow:

Total time of tracking scenario is 40sec. Initial target position is x=-2000m, y=500m and z=700m. In the first part for 5 seconds, target has constant velocity $v$=200 m/sec along the x axis. In the second part for 7 seconds, target turns with constant turn rate 18deg/sec and constant velocity 200m/sec around z axis. In the third part for 5 seconds, target has constant velocity 200m/sec. In the fourth part, for 8 seconds target turns with constant turn rate -22.5deg/sec and constant velocity 200m/sec around z axis. In the fifth part, for 5 seconds target has constant acceleration 0.3m/sec$^2$ along body x axis and in the sixth part, for 10 seconds target has constant jerk 10m/sec$^3$ along body x axis. During entire maneuverer, target has constant velocity 50m/sec along the z axis. It is assumed that optical frame sampling interval ($T_s$) is 33milliseconds. Measurement noise in $\theta$, $\psi$ and $r$ is assumed to be white Gaussian noise with standard deviations $\sigma_\theta$=0.02deg, $\sigma_\psi$=0.02deg and $\sigma_r$=3m respectively.

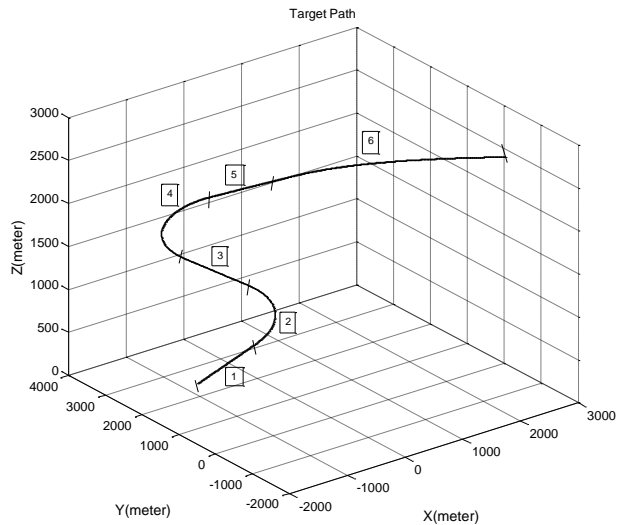

**Fig. 2** Path of a maneuvering target in simulated tracking scenario. Six phases of target maneuver are specified.

Fig. 2 shows target path in the space and Fig. 3 shows its related x, y and z components. In Figure 4, for given tracking scenario and with range measurement scheduling, target state estimation errors and their $3\sigma$ bounds for IMM-UKF based estimator are demonstrated. The states whose estimation errors are shown are $\omega$, $\dot{\theta}$, $\tau$, $\psi$, $\theta$ and $s$. It can be seen that target state estimation errors for all states except for $\omega$, are approximately in $3\sigma$ bounds. For state $\omega$, in time duration between 30 and 40, NCA model does not match with target constant jerk maneuver and therefore estimation error is out of $3\sigma$ bounds. Hence IMM-UKF filter is reasonably consistent. Fig. 5 shows estimated target range, range estimation error, range scheduling pulses and Root Mean Square (RMS) value of target range estimation error for Monte-Carlo simulation with 100 runs using IMM-UKF based tracker in MSC for given tracking scenario. For proper filter initialization, in the first fifty frame of scenario time (t=0 to t=1.65sec.), regular range measurement is considered and after that, range measurements are scheduled. It can be seen that during target maneuver, the rate of range measurement scheduling pules is higher than the

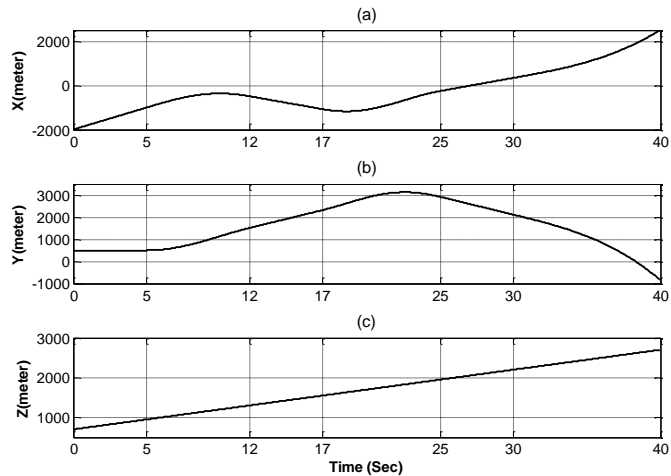

**Fig. 3** x, y and z component of target position in Cartesian coordinate frame; (a) x component, (b) y component, (c) z component. Different target maneuver phases are separated by grid lines on time axis .

periods with no target maneuver. In Figures 6, mixing probabilities for IMM-UKF tracking filter are demonstrated. This figure shows that the proposed tracking filter can detect target maneuvers correctly. Finally Fig. 7 shows real and estimated target turn rate and acceleration. It can be seen that target acceleration and turn rate are estimated correctly.

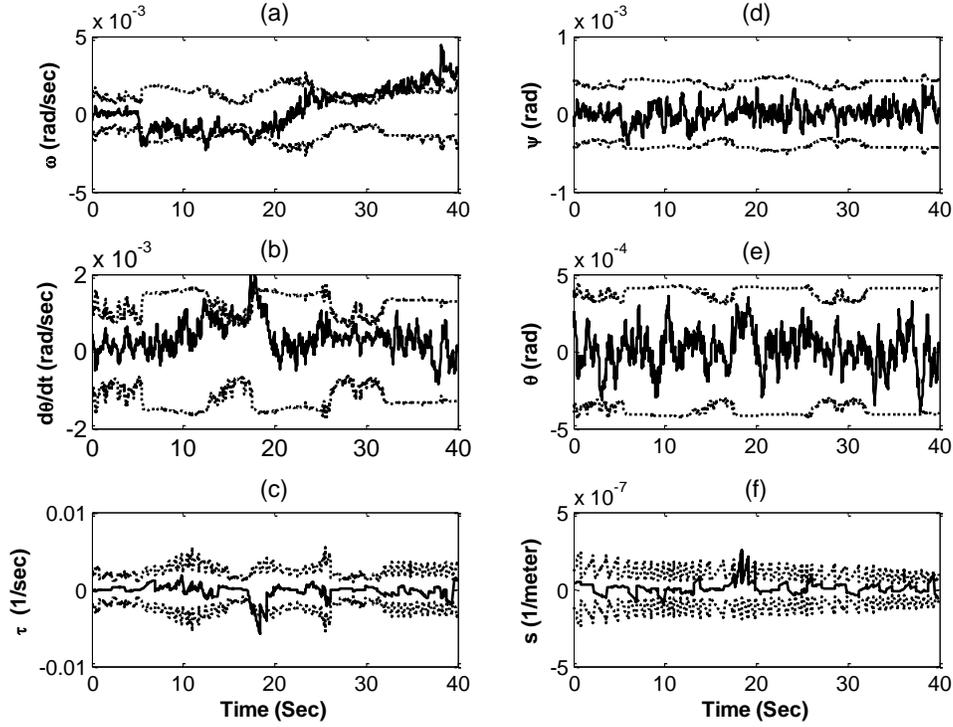

**Fig. 4** Target state estimation errors and related $3\sigma$ bounds for proposed IMM-UKF based tracker with the given tracking scenario and with range measurement scheduling; (a) $\omega$, (b) $\frac{d\theta}{dt}$, (c) $\tau$, (d) $\psi$, (e) $\theta$, (f) $s$. Solid line shows the estimated states and dotted lines show three-sigma limits for estimated states.

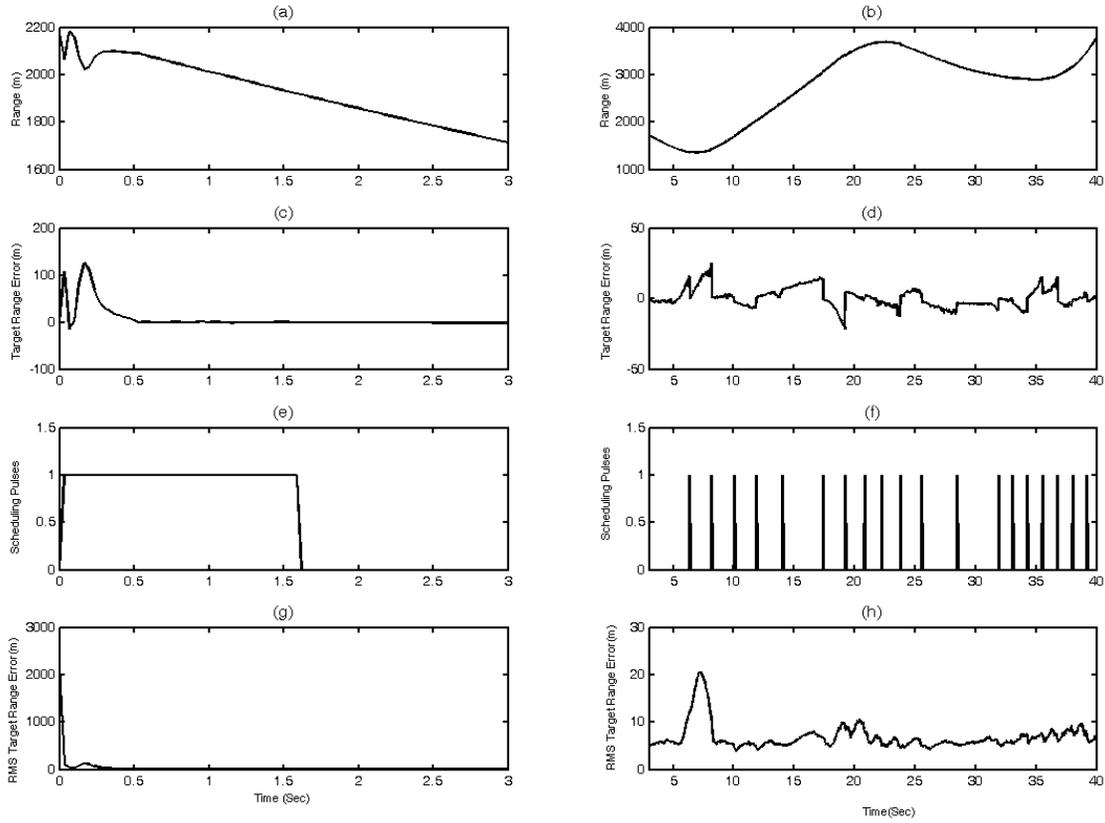

**Fig. 5** Simulation results for the proposed IMM-UKF based tracking filter with the given tracking scenario; in the first column, simulation results from t=0 to 3 sec and in the second column, simulation results for t=3 sec to the end of scenario time are shown. Note that in the first fifty frame of scenario time (t=0 to t=1.65 sec), regular target range measurement (without scheduling) is performed. (a,b) Estimated range, (c,d) Range estimation error, (e,f) Range measurement scheduling pulses, (h ,g) RMS of range estimation error with 100 Monte-Carlo run.

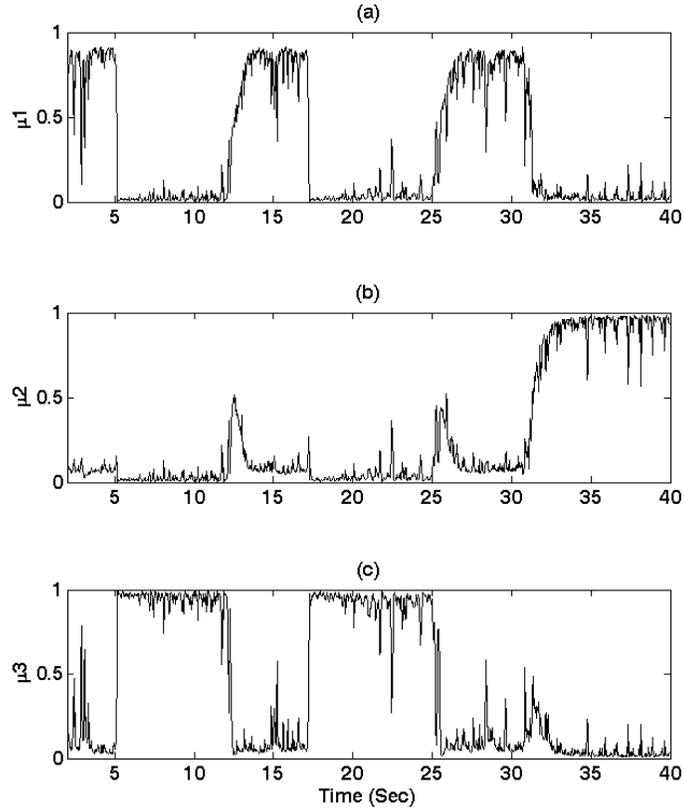

**Fig. 6** Mixing probabilities for proposed IMM-UKF based tracker with the given tracking scenario; (a) µ1 mixing probability for NCV model, (b) µ2 mixing probability for NCA model, (c) µ3 mixing probability for CT model.

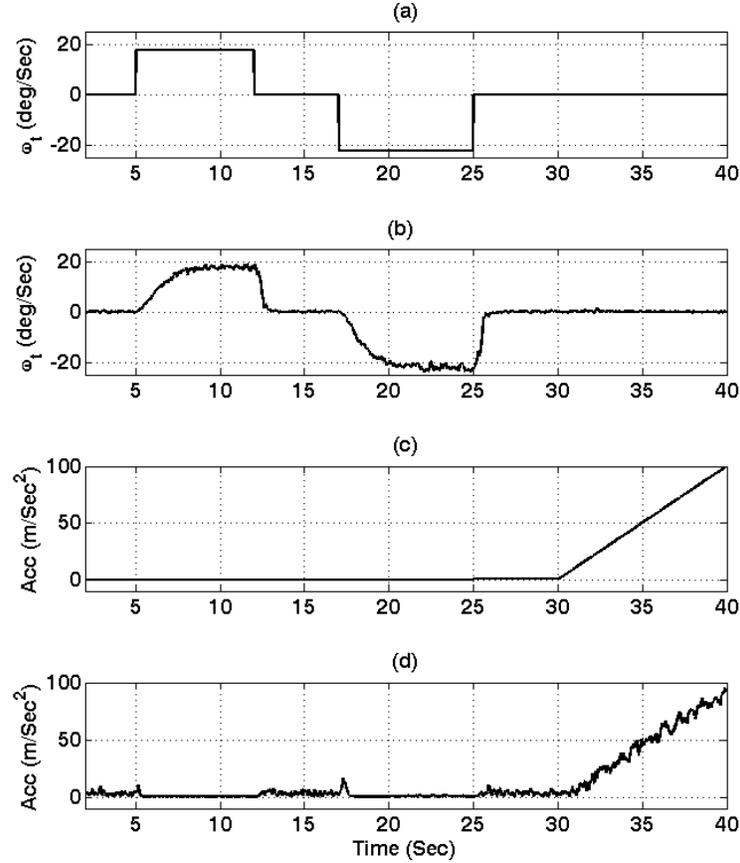

**Fig. 7** Comparison of real and estimated target turn rate and acceleration for the proposed IMM-UKF based tracker with the given tracking scenario; (a) Actual target turn rate, (b) Estimated target turn rate, (c) Actual target acceleration, (d) Estimated target acceleration.

## 8 Conclusion

In this paper, for optical target tracking by scheduled range measurements, a new set of target states in MSC have been presented. Proposed set of target states includes inverse target range and this state does not appear in dynamic equations of the other states, therefore an appropriate situation for range measurement scheduling is achieved. For improving tracking performance in maneuvering target tracking scenarios, IMM based tracking filter with NCV, NCA and CT kinematic models have been considered. Moreover dynamic equations of target states for these kinematic models have been derived in MSC. In order to have scheduled range measurements, standard deviation of range estimation error has been utilized as a scheduling criterion. Active range measurement is performed if scheduling criterion exceeds a predetermined threshold. Simulation results show that the proposed IMM-UKF based tracker has appropriate performance in maneuvering target tracking and range measurement scheduling.

**Mohammad Hossein Ferdowsi** was born in Esfahan, Iran, in 1953. He received the B.Sc. and M.Sc. degrees in Electrical Engineering from Sharif University of Technology and Ph.D. degree in Electrical Engineering from University of Tehran in 1977, 1980 and 2004 respectively. He has worked in industry for five years as an Instrumentation and Control Engineer. From 1985 to 1987 he has been a lecturer at Sharif University of technology and from 1987 up to now he has been at Malek-e-Ashtar University of Technology as a faculty member. His research interest is in multivariable and adaptive control systems, intelligent systems and target tracking.

**Ebrahim Sabzikar** was born in Qazvin, Iran in 1984. He received B.Sc. in Electrical Engineering from Iran University of Science and Technology (IUST), and M.Sc. in Control Engineering form Amir-Kabir University (Tehran Polytechnic), Tehran, Iran respectively in 2007 and 2011. He is currently pursuing the Ph.D. degree at Malek-e-Ashtar University of Technology, Tehran, Iran. His interested researches include signal processing, tracking, and data fusion.

**Appendix A: Derivation of state dynamic equations for NCV kinematic model**

Spherical coordinate frame is a non-inertial frame; therefore Newton's second law is not hold. For analytical derivation of state dynamic equations in MSC applying Newton's second law, transport theorem in Ref. 17 is utilized.

**Transport Theorem**: If *B* and *N*, be two frames with a relative angular velocity vector of $\boldsymbol{\omega}_{NB}$, and let $\boldsymbol{r}$ be a generic vector; then the derivative of $\boldsymbol{r}$ in *N* frame can be related to derivative of $\boldsymbol{r}$ in the *B* frame as

$$\frac{^N d}{dt}(\boldsymbol{r}) = \frac{^B d}{dt}(\boldsymbol{r}) + \boldsymbol{\omega}_{NB} \times \boldsymbol{r}, \tag{A-1}$$

In (A-1). sign ($\times$) shows cross product. If $\boldsymbol{r}^N$ be representation of generic vector $\boldsymbol{r}$ in *N* frame, then $\frac{^N d}{dt}\boldsymbol{r}^N$ will be written as a compact notation $\dot{\boldsymbol{r}}^N$.

Cartesian and spherical coordinate frames are represented using superscript *c* and *s* respectively and it is assumed that Cartesian coordinate frame is an inertial frame. Applying Newton's second law to inertial frame *c*, acceleration $\boldsymbol{a}^c$ is resulted as (A-2).

$$\boldsymbol{a}^c = \frac{^c d}{dt}\left(\frac{^c d}{dt}\boldsymbol{r}^c\right) = \frac{^c d}{dt}\dot{\boldsymbol{r}}^c \tag{A-2}$$

For spherical coordinate frame, acceleration $\boldsymbol{a}^s$ is as (A-3).

$$\boldsymbol{a}^s = \frac{^c d}{dt}\left(\frac{^c d}{dt}\boldsymbol{r}^s\right) \tag{A-3}$$

For extracting nonlinear dynamic equations of target states in MSC, it is assumed that the vector $\boldsymbol{r}^s$ is defined as (A-4).

$$\boldsymbol{r}^s = [r \quad 0 \quad 0]^T, \tag{A-4}$$

Time derivative for this vector in Cartesian coordinate frame using transport theorem is given by (A-5).

$$\frac{^c d}{dt}r^s = \frac{^s d}{dt}r^s + \omega^s_{cs} \times r^s = \dot{r}^s + \omega^s_{cs} \times r^s, \tag{A-5}$$

In (A-5), $\omega^s_{cs}$ is the relative angular velocity of spherical coordinate frame with respect to Cartesian coordinate frame stated in spherical coordinate frame. Second derivative of the vector $r^s$ with respect to Cartesian coordinate frame is defined by

$$\frac{^c d^2}{dt^2}r^s \triangleq \frac{^c d}{dt}\left(\frac{^c d}{dt}r^s\right) = \ddot{r}^s + 2\omega^s_{cs} \times \dot{r}^s + \dot{\omega}^s_{cs} \times r^s + \omega^s_{cs} \times \omega^s_{cs} \times r^s = C^s_c a^c, \tag{A-6}$$

In (A-6), $C^s_c$ is rotation matrix from Cartesian to spherical frame and is defined as (A-7).

$$C^s_c = C_y(-\theta)C_z(\psi), \tag{A-7}$$

$$C_y(\theta) = \begin{bmatrix} \cos\theta & 0 & -\sin\theta \\ 0 & 1 & 0 \\ \sin\theta & 0 & \cos\theta \end{bmatrix}, \tag{A-8}$$

$$C_z(\psi) = \begin{bmatrix} \cos\psi & \sin\psi & 0 \\ -\sin\psi & \cos\psi & 0 \\ 0 & 0 & 1 \end{bmatrix}, \tag{A-9}$$

Also, $a^c$ is target acceleration in Cartesian coordinate frame that is considered to be white noise for nearly constant velocity (NCV) model given as (A-10).

$$a^c = w^c = [a^c_x \quad a^c_y \quad a^c_z]^T = [w^c_x \quad w^c_y \quad w^c_z]^T, \tag{A-10}$$

$$a^s = w^s = C^s_c w^c = [w^s_x \quad w^s_y \quad w^s_z]^T = [a^s_x \quad a^s_y \quad a^s_z]^T, \tag{A-11}$$

Vector $\omega^s_{cs}$ and its derivative are as (A-12) and (A-13).

$$\omega^s_{cs} = [\dot{\psi}\sin\theta \quad -\dot{\theta} \quad \dot{\psi}\cos\theta]^T, \tag{A-12}$$

$$\dot{\omega}^s_{cs} = [\ddot{\psi}\sin\theta + \dot{\psi}\dot{\theta}\cos\theta \quad -\ddot{\theta} \quad \ddot{\psi}\cos\theta - \dot{\psi}\dot{\theta}\sin\theta]^T, \tag{A-13}$$

Furthermore, $\dot{r}^s$ and $\ddot{r}^s$ are defined as (A-14).

$$\dot{r}^s = [\dot{r} \quad 0 \quad 0]^T, \ddot{r}^s = [\ddot{r} \quad 0 \quad 0]^T, \tag{A-14}$$

Expanding equation (A-6) generates three new equations as (A-15), (A-16) and (A-17).

$$w_x^s = \ddot{r} - r\dot{\theta}^2 - r\dot{\psi}^2\cos^2\theta, \tag{A-15}$$

$$w_y^s = 2\dot{r}\dot{\psi}\cos\theta + r\ddot{\psi}\cos\theta - 2r\dot{\psi}\dot{\theta}\sin\theta, \tag{A-16}$$

$$w_z^s = r\sin\theta\cos\theta\dot{\psi}^2 + 2\dot{r}\dot{\theta} + r\ddot{\theta}, \tag{A-17}$$

After some manipulations, equations (A-18), (A-19) and (A-20) are resulted.

$$\ddot{r} = r(\dot{\theta}^2 + (\dot{\psi}\cos\theta)^2) + w_x^s, \tag{A-18}$$

$$\frac{d}{dt}(\dot{\psi}\cos\theta) = -2\frac{\dot{r}}{r}\dot{\psi}\cos\theta + \dot{\psi}\dot{\theta}\sin\theta + \frac{w_y^s}{r}, \tag{A-19}$$

$$\ddot{\theta} = -2\frac{\dot{r}}{r}\dot{\theta} - \dot{\psi}^2\sin\theta\cos\theta + \frac{w_z^s}{r}, \tag{A-20}$$

Now variables in spherical coordinate frame are written with respect to MSC in (A-21) and (A-22).

$$\ddot{r} = \frac{\dot{\tau} + \tau^2}{s}, \tag{A-21}$$

$$\omega = \dot{\psi}\cos\theta, \tag{A-22}$$

Therefore state dynamic equations in MSC are given a (3).

**Appendix B: Derivation of state dynamic equations for NCA kinematic model**

In (3), assuming that $a_x^c$, $a_y^c$ and $a_z^c$ are not white Gaussian random process, these parameters can be modeled as a first order wiener random process as (B-1).

$$\dot{a}_x^c = w_x^c, \dot{a}_y^c = w_y^c, \dot{a}_z^c = w_z^c, \tag{B-1}$$

Furthermore, expanding equation (A-11) generates new equations as (B-2).

$$\begin{cases} a_x^s = \cos\theta\,\cos\psi\,a_x^c + \cos\theta\,\sin\psi\,a_y^c + \sin\theta a_z^c \\ a_y^s = -\sin\psi\,a_x^c + \cos\psi a_y^c \\ a_z^s = -\sin\theta\cos\psi\,a_x^c - \sin\theta\sin\psi a_y^c + \cos\theta\,a_z^c \end{cases}, \tag{B-2}$$

By replacing (B-1) and (B-2) in (3), new equations obtained as (B-3).

$$\begin{cases} \frac{d\omega}{dt} = -2\tau\omega + \dot{\theta}\omega \tan\theta - \sin\psi \, s \, a_x^c + \cos\psi \, s \, a_y^c \\ \frac{d\dot{\theta}}{dt} = -\omega^2 \tan\theta - 2\dot{\theta}\tau - \sin\theta \cos\psi \, s \, a_x^c - \sin\theta \sin\psi \, s \, a_y^c + \cos\theta \, s \, a_z^c \\ \frac{d\tau}{dt} = \dot{\theta}^2 + \omega^2 - \tau^2 + \cos\theta \cos\psi \, s \, a_x^c + \cos\theta \sin\psi \, s \, a_y^c + \sin\theta \, s \, a_z^c \\ \frac{da_x^c}{dt} = w_x \\ \frac{da_y^c}{dt} = w_y \\ \frac{da_z^c}{dt} = w_z \end{cases} \quad \text{(B-3)}$$

It can be seen that state $s$ is observed in dynamic equations of the other states. In order to eliminate dependence on $s$ in the other state dynamic equations, three new states $\sigma_x^c$, $\sigma_y^c$ and $\sigma_z^c$ are defined as (B-4).

$$\sigma_x^c = s \, a_x^c, \; \sigma_y^c = s \, a_y^c, \; \sigma_z^c = s \, a_z^c, \quad \text{(B-4)}$$

Derivative of state $s$ is defined as (B-5).

$$\frac{ds}{dt} = -\tau \, s, \quad \text{(B-5)}$$

Using equations (B-1) and (B-5) and time derivative of terms for both sides of equalities in (B-4), dynamic equations of these new states are obtained as (B-6).

$$\begin{cases} \frac{d\sigma_x^c}{dt} = -\tau \, \sigma_x^c + s \, w_x \\ \frac{d\sigma_y^c}{dt} = -\tau \, \sigma_y^c + s \, w_y, \\ \frac{d\sigma_z^c}{dt} = -\tau \, \sigma_z^c + s \, w_z \end{cases} \quad \text{(B-6)}$$

In dynamic equations (B-3), $w_x$, $w_y$ and $w_z$ are jerk process noise and state $s$ is treated as a process noise gain in the state dynamic equations. Dynamic equations of these states are as (5).

**Appendix C: Derivation of state dynamic equations for CT kinematic model**

Second derivative of the vector $r^s$ with respect to Cartesian coordinate frame is defined as (A-6). In (A-6) and for CT model, $a^c$ is target acceleration vector in Cartesian coordinate frame and assuming that target coordinated turn maneuver is in x-y plane, it is defined as (C-1).

$$a^c = w^c = [a_x^c \quad a_y^c \quad a_z^c]^T = [w_x^c \quad w_y^c \quad w_z^c]^T + [-\omega_T v_y^c \quad \omega_T v_x^c \quad 0]^T, \tag{C-1}$$

In (C-1), $\omega_T$ is the turn rate of the target, $v_x^c$ and $v_y^c$ are target velocities in x and y axis of Cartesian coordinate frame. Also $w_x^c$, $w_y^c$ and $w_z^c$ are acceleration process noises in Cartesian coordinate frame. $v_x^c$ and $v_y^c$ can be stated in MSC as (C-2).

$$\begin{cases} v_x^c = \dot{r}\cos(\theta)\cos(\psi) - r\dot{\theta}\sin(\theta)\cos(\psi) - r\dot{\psi}\cos(\theta)\sin(\psi) \\ v_y^c = \dot{r}\cos(\theta)\sin(\psi) - r\dot{\theta}\sin(\theta)\sin(\psi) + r\dot{\psi}\cos(\theta)\cos(\psi) \end{cases}, \tag{C-2}$$

$w^s$ is acceleration process noise vector in spherical coordinate frame as (C-3).

$$w^s = [w_x^s \quad w_y^s \quad w_z^s]^T = C_c^s [w_x^c \quad w_y^c \quad w_z^c]^T, \tag{C-3}$$

By replacing (C-1), (C-2), (C-3) and (A-11) in (A-6) and some manipulations, three new equations are obtained as (C-4), (C-5) and (C-6).

$$\ddot{r} = r(\dot{\theta}^2 + (\dot{\psi}cos\theta)^2 - \omega_T \dot{\psi}\cos^2\theta) + w_x^s, \tag{C-4}$$

$$\frac{d}{dt}(\dot{\psi}cos\theta) = -2\frac{\dot{r}}{r}\dot{\psi}cos\theta + \dot{\psi}\dot{\theta}sin\theta - \dot{\theta}\omega_T \sin\theta + \tau\omega_T \cos\theta + \frac{w_y^s}{r}, \tag{C-5}$$

$$\ddot{\theta} = -2\frac{\dot{r}}{r}\dot{\theta} - \dot{\psi}^2 sin\theta cos\theta + \omega_T \sin\theta \, \omega + \frac{w_z^s}{r}, \tag{C-6}$$

Using (A-16) and (A-17), target state dynamic equations for CT model in MSC are given as (7).

29